\documentclass[a4paper]{jpconf}
\usepackage{graphicx}

\newcommand{ \be }{\begin{equation}}
\newcommand{ \ee }{\end{equation}}
\newcommand{ \bea }{\begin{eqnarray}}
\newcommand{ \eea }{\end{eqnarray}}

\usepackage{color}

\begin{document}
\title{Hadron production in ultra-relativistic nuclear collisions and the QCD
  phase boundary.} 

\author{Peter Braun-Munzinger        }
\address{GSI and Technical University,\\ 
        Darmstadt, \\ 
        Germany}

\begin{abstract}
We update briefly our understanding of  hadron production in relativistic
nucleus-nucleus  collisions in terms of statistical  models with emphasis on
the relation of the data to the QCD phase boundary and on a puzzle in the beam
energy dependence.   

\end{abstract}

\section{Hadron yields, statistical description, and the phase boundary to the
  QGP.}

\begin{figure}[hbt]
\begin{center}
\vspace{-.50cm}
\includegraphics[width=16.0cm]{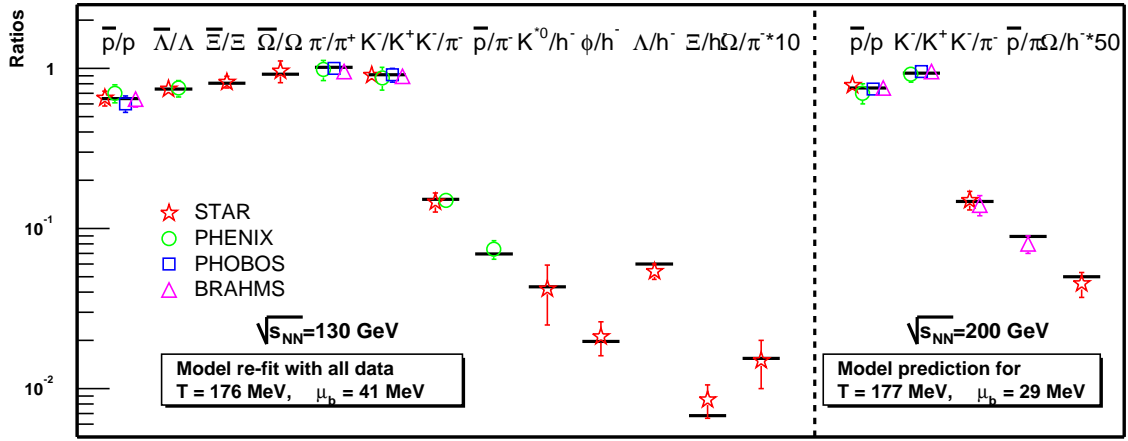} 
\vspace{-.80cm}
\caption{Fit of particle ratios for Au-Au collisions measured at RHIC energies.
The measurements are the symbols, the thermal model values are the lines
   \cite{rhic,review}}
\label{yields}
\end{center}
\end{figure}

Hadron yields or more specifically yield ratios observed in central nuclear
collisions at AGS, SPS, and RHIC energies can be described with high precision
within a hadro-chemical equilibrium approach
\cite{agssps,satz,heppe,cley,beca1,rhic,nu,beca2,rapp}, governed by a chemical
freeze-out temperature T$_{ch}$ and baryo-chemical potential $\mu$. A recent
review, found in \cite{review}, provides a wealth of information on the
subject.  The data at SPS and RHIC energy comprise multi-strange hadrons
including the $\Xi$, $\bar \Xi$, $\Omega$ and $\bar \Omega$ baryons. Their
yields ratios (to pions, e.g.)  agree particularly well with the chemical
equilibrium calculation and are enhanced by more than an order of magnitude as
compared to observations in pp collisions. We present as an example the results
for central Au-Au collisions at RHIC energy in Fig.~\ref{yields}.

A remarkable exception is observed for strongly decaying, wide resonances.  In
particular, the $\Delta/p$ ratio is overpredicted while the $\rho/\pi$ ratio,
both determined in (rather peripheral, though) Au-Au collisions by the STAR
collaboration \cite{star_wide1,star_wide2}, is underpredicted  by the model
\cite{review}. For an interesting and provocative speculation as to where this
might come from see \cite{gerry}. Clearly precision data for central
collisions and also at other beam energies are needed to settle this issue.

\begin{figure}[hbt]
\begin{center}
\vspace{-1.3cm}
\includegraphics[width=12.0cm]{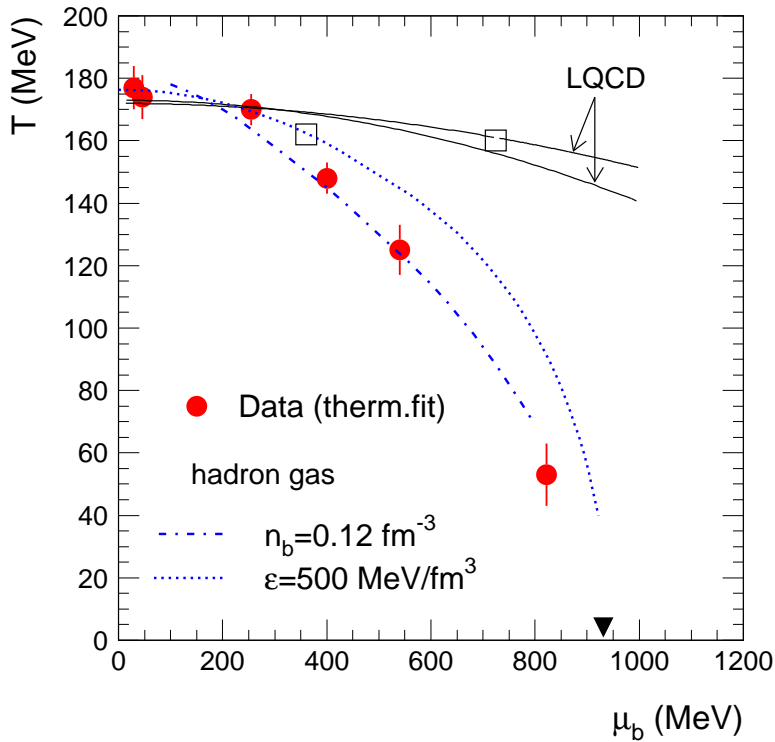} 
\vspace{-.80cm}
\caption{Phase diagram of hadronic matter and chemical freeze-out points
  \cite{andronic1}. The 
  open squares represents  recent estimates \cite{karsch2,fodor} of the
  position of the tri-critical point (see
  text). 
}
\label{phase}
\end{center}
\end{figure}

The chemical parameters T$_{ch}$ and $\mu$ determined from fits to data at all
available energies are plotted in the QCD phase diagram shown in
Fig.~\ref{phase}, taken (slightly modified) from \cite{andronic1}. The phase
transition lines in this figure are obtained from recent analyses within the
framework of lattice QCD \cite{karsch2,fodor} and include recent estimates
\cite{karsch2,fodor} of the position of the tri-critical
point\footnote{Position and even the existence of such a tri-critical point
  are currently hotly debated.  For a critical discussion see
  \cite{philipsen}.}. Also included in this figure are lines of constant total
baryon density and lines of constant energy density, computed within the
framework of the hadronic gas model of \cite{heppe}.  The line of constant
energy density $\epsilon = 500 $ MeV/fm$^3$ should in our view be a reasonable
phenomenological description of the phase boundary. Interestingly, it is, at
$\mu > 500$ MeV, much closer to the chemical freeze-out points than to the
phase lines from lattice QCD. Apparently, the phase boundary lines obtained
from state-of-the-art calculations within the framework of lattice QCD are not
at all lines of constant energy density.  Rather, the energy density grows by
more than a factor of 2 when going from 0 to 500 MeV in $\mu$. This rather
surprizing result deserves further attention.

An important observation, already made in\cite{agssps}, concerning the results
of the thermal model calculations is that, for top SPS energy and above, the
chemical parameters determined from the measured hadron yields coincide within
the uncertainties of about $\pm 10$ MeV with the phase boundary as determined
from lattice QCD calculations.  A natural question arises: is this coincidence
accidental and, if not, what enforces equilibration at the phase boundary?
Considerations about collisional rates and timescales of the hadronic fireball
expansion \cite{wetterich} imply that at SPS and RHIC energies the equilibrium
cannot be established in a purely hadronic medium and that is is the phase
transition which drives the particles densities and ensures chemical
equilibrium.

In \cite{wetterich} it is further shown that many body collisions become
important close to T$_c$ and provide a mechanism for rapid equilibration.
Because of the strong density increase near the QCD phase transition such
multi-particle collisions provide a natural explanation for the observation of
chemical equilibration at RHIC energies and lead to T$_{ch}$=T$_c$ within an
accuracy of a few MeV. Any scenario with T$_{ch}$ substantially smaller than
T$_c$ would require that either multi-particle interactions dominate even much
below T$_c$ or that the two-particle cross sections are larger than in the
vacuum by a high factor. Both of the latter hypotheses seem unlikely in view
of the rapid density decrease in the hadronic phase of the expanding fireball.
The critical temperature determined from RHIC for T$_{ch}\approx \rm{T_c}$
coincides well with recent lattice estimates \cite{karsch} for $\mu=0$.  The
same arguments as discussed here for RHIC energy also hold for SPS energies:
it is likely that also there the phase transition drives the particle
densities and insures chemical equilibration.

\begin{figure}[hbt]
\begin{center}
\vspace{-1.50cm}
\includegraphics[width=12.0cm]{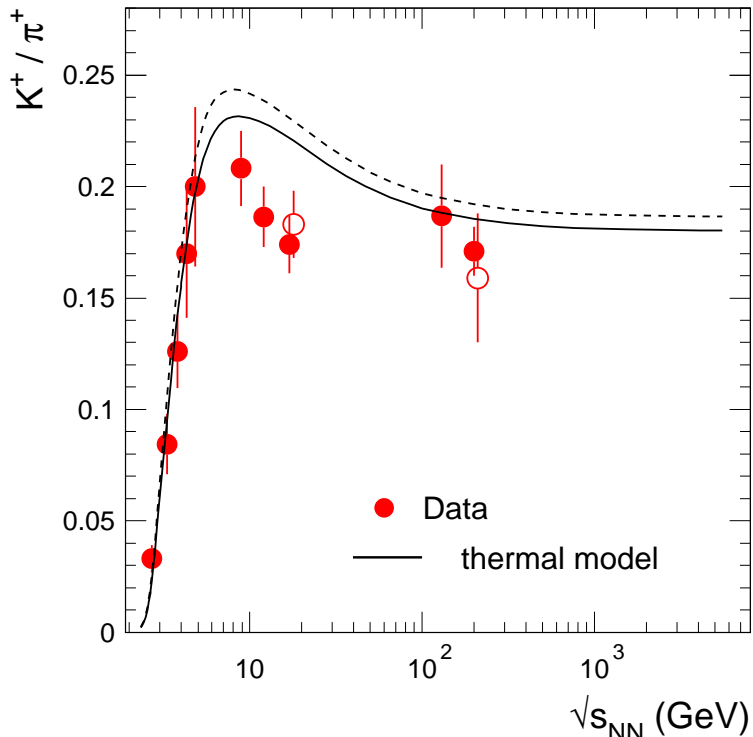} 
\vspace{-.80cm}
\caption{The K$^+/\pi^+$ ratio from NA49 \cite{marek} and thermal model
  predictions from \cite{andronic_new}. The dashed and solid lines correspond
  to calculations without and with account of the resonance widths.
}
\label{k2pi}
\end{center}
\end{figure}

It was alternatively proposed \cite{stock,heinz2,heinz1} that the observed
hadron 
abundances arise from a direct production of strange (and non-strange)
particles by hadronization. How this happens microscopically is
unclear.  To escape the above  argument that
T$_{ch}$=T$_c$ one would have to argue that the particle yields are
established without hadronic rescattering. This is unlikely since the
abundances are determined by hadronic properties (masses) with high
precision. Second, one may question if the ``chemical temperature''
extracted from the abundances is a universal temperature which also
governs the local kinetic aspects and can be associated with the
critical temperature of a phase transition in equilibrium. Indeed, in
a prethermalization process, different equilibrium properties are
realized at different time scales. Nevertheless, all experience shows
that kinetic equilibration occurs before chemical equilibration. It
seems hard to imagine that chemical equilibrium abundances are
realized at a time when the kinematic distributions
are not yet close to their equilibrium
values.

\begin{figure}[hbt]
\begin{center}
\vspace{-1.50cm}
\includegraphics[width=12.0cm]{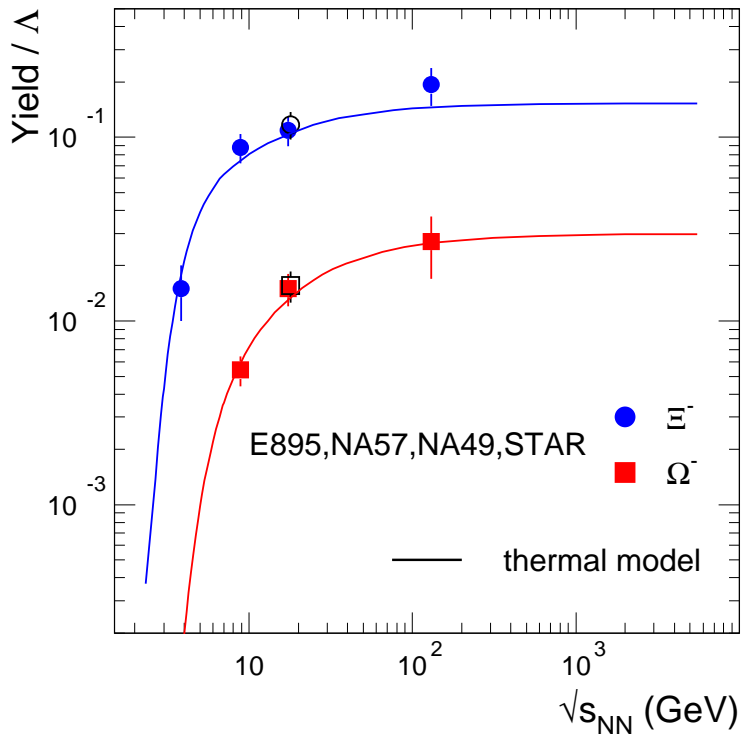} 
\vspace{-.80cm}
\caption{Energy dependence of particle rations involving strange baryons
  compared to thermal model calculations of \cite{andronic_new}.
}
\label{xiom}
\end{center}
\end{figure}

The chemical equilibrium curve and the QCD phase boundary as obtained from
lattice QCD calculations begin to differ for baryo-chemical potential values
of $\mu > $ 400 MeV. Despite that multi-strange baryons are produced with
yields very close to what is obtained in a full chemical equilibration
scenario (see below). This could be an indication for new, as yet unidentified
equilibration
processes, different from those advocated in \cite{wetterich} and also
incompatible with the ``phase space filling'' scenario through hadronization
put forward by \cite{stock,heinz1,heinz2}, in a dense, baryon-rich
medium. Considering the discussion above, a rather radical but appealing
possibility is that the chemical freeze-out curve defines the phase boundary
also at large $\mu$ values up to $\mu \approx 600$ MeV. We note in this
context that the chemical freeze-out points determined from measurements at
SIS energy should still be considered with caution as no multi-strange baryons
have been measured there.

The relatively smooth energy dependence of the above discussed thermal
parameters is to be contrasted with recent observation by the NA49
collaboration  of an anomaly in the K$^+/\pi^+$ ratio
near $\sqrt{s_{nn}} = 8$ GeV \cite{marek,blume}.  These results look rather
striking, and indeed, the narrowness of the observed structure is missed by
the thermal model predictions, as is visible in Fig.~\ref{k2pi}.

We note, however, that the anomaly seems to be confined to ratios involving
light mesons. For strange baryons, the situation is quite different, as is
demonstrated in Fig.~\ref{xiom}. Here, the observed energy dependence is very
well reproduced by the calculations and no anomaly is visible.

The situation is even more puzzling considering that the energy averaging
inherent in ultra-relativistic nucleus-nucleus collisions leads to a
characteristic smearing of all possible intrinsic structures. The effect can
be estimated \cite{edep} using simple geometrical arguments and is of the
order 10 \% of the cm. energy per nucleon at SPS energy. The corresponding
width is very close to the width of the structure observed by the NA49
collaboration and would imply that the strength of the intrinsic structure is
even much larger than visible in Fig.~\ref{k2pi}. Further research is
necessary to shed light on this puzzle.

{\it Acknowledgments.}
Discussions with Anton Andronic and 
Johanna Stachel are gratefully acknowledged.

\end{document}